# Efficient Simulation of Temperature Evolution of Overhead Transmission Lines Based on Analytical Solution and NWP

Rui Yao, *Member, IEEE,* Kai Sun, *Senior Member, IEEE,* Feng Liu, *Member, IEEE,* and Shengwei Mei, *Fellow, IEEE*

*Abstract* --- **Transmission lines are vital components in power systems. Outages of transmission lines caused by over-temperature is a major threat to system reliability, so it is necessary to efficiently simulate line temperature under both normal operation conditions and foreseen fault conditions. Existing methods based on thermal-steady-state analyses cannot support the simulation of transient temperature evolution, and thus cannot provide timing information needed for taking remedial actions. Moreover, conventional numerical method requires huge computational efforts and barricades system-wide analysis. In this regard, this paper derives an approximate analytical solution of transmission-line temperature evolution enabling efficient analysis on multiple operation states. Considering the uncertainties in environmental parameters, the region of over-temperature is constructed in the environmental parameter space to realize the over-temperature risk assessment in both the planning stage and real-time operations. A test on a typical conductor model verifies the accuracy of the approximate analytical solution. Based on the analytical solution and numerical weather prediction (NWP) data, an efficient simulation method for temperature evolution of transmission systems under multiple operation states is proposed. As demonstrated on a Northeast Power Coordinating Council (NPCC) 140-bus system, it achieves over 1000 times of efficiency enhancement, verifying its potential in online risk assessment and decision support.**

## I. INTRODUCTION

TRANSMISSION lines play vital roles in conveying electricity from the plants to the users in the power systems [1]. Also, overhead transmission lines cover largest area across the system, and they are exposed to complex environmental conditions [2]. Many kinds of natural events threaten the transmission lines, such as lightning, strong wind, hot weather, etc. [3, 4]. A transmission line has certain thermal capacity, and once the temperature exceeds its limit, the line may face risk of tripping due to sagging and tree contact [5, 6] or outage due to heat damage [7]. Such threats are particularly prominent on heavily-loaded lines exposed to high ambient temperature and low wind conditions. The outage of transmission lines due to over-temperature is a major cause of cascading outages in power systems (e.g. the 1996 WSCC outages [8] and 2003 US-Canada blackout [9]). Moreover, the fast-growing load and developing electricity market but relatively slow upgrade of transmission infrastructure has pushed transmission lines toward operational limits [10]. Therefore, it is necessary to enhance the situational awareness of transmission lines under various environmental conditions [11]. Specifically, it is desirable to realize accurate and efficient simulation of future transmission-line temperature evolution (TLTE) by considering environmental factors.

To exploit the transmission capacity and to monitor the lines, the dynamic line rating (DLR) was proposed to determine the maximum current flow under which the steady-state temperature will not exceed the limit [12]. However, the DLR only studies steady state temperature. In fact, it is also important to address the temperature transients to obtain more accurate and panoramic information on the risk of the transmission system over a timespan. Besides normal operating conditions, it is also necessary to study other operation states, e.g. when the system is under contingencies or cascading failures [9]. This is particularly useful and adds to system robustness in case some elements are under maintenance, or operators are unaware of loss of components due to malfunctions in the SCADA/EMS [9], or when the system is under cyber-attacks [13]. IEEE [14] and CIGRE [15] have proposed models describing TLTE. An existing approach to simulating the TLTE is numerical integration on the differential equation. However, such an approach requires huge computational efforts and is computationally difficult in system-wide analysis because: 1) a line should be divided and simulated in many segments since the environmental parameters are different along the line; 2) when system state changes, a complete system-wide simulation is required.

To overcome the limitations of conventional numerical methods, this paper first derives an approximate analytical solution of TLTE. The proposed approximate analytical solution can significantly improve the efficiency of simulation and enhance the practicality for the monitoring and analysis of transmission lines in four folds:

1) The approximate analytical solution significantly improves the efficiency of simulating TLTE by simply assigning values to the variables in the analytical expression, which avoids cumbersome numerical integration.

2) This paper proposes a method for quickly updating the analytical solution when current changes, which further enhances computational speed in batch analysis of multiple operation states (e.g. analysis of a large set of contingencies).

This work was supported by the CURENT Engineering Research Center.

R. Yao is with State Key Laboratory of Power Systems, Department of Electrical Engineering, Tsinghua University, Beijing 100084, China, and the Department of EECS, University of Tennessee, Knoxville, TN 37996, USA. (email: ryao3@utk.edu, yaorui.thu@gmail.com)

K. Sun is with the Department of EECS, University of Tennessee, Knoxville, TN 37996, USA (email: kaisun@utk.edu).

F. Liu and S. Mei are with the State Key Laboratory of Power Systems, Department of Electrical Engineering, Tsinghua University, Beijing 100084, China. (email: lfeng@mail.tsinghua.edu.cn, meishengwei@mail.tsinghua.edu.cn).

3) Useful metrics for the security analysis and decision-making can be derived from the analytical solution, e.g. the time for a conductor to reach the temperature limit.

4) The analytical solution can be further utilized in more advanced risk modeling, analysis and optimization for transmission systems [16].

Moreover, this paper uses the high resolution numerical weather prediction (NWP) service as the source of environmental data. The NWP has become a mature public service providing with reliable wide-area, high-resolution environmental data, which also has potential for applications in power systems [17]. In the US, the NWP previously only provides an hourly forecast, which is too coarse to meet the temporal resolution for the simulation of transmission line temperature [18]. Recently a new model of NWP in North America -- the High-Resolution Rapid Refresh Version 2 (HRRR-v2) model has been put into operations starting from August 2016, providing an hourly-refreshed forecast at the 15-minute interval and 3km spatial resolution covering contiguous US (CONUS) and Alaska with outreach of 18 hours [19]. Thus, the NWP turns out to be a promising source of environmental data for assessment and monitoring of real-time reliability of transmission lines.

The areas with low wind and high ambient temperature have high risk of over-temperature occurrence. Although the NWP models have considered influence of terrain on the near-surface airflow, they cannot address the terrain in scales smaller than the NWP spatial resolution. However, the smaller-scale terrain may also have substantial impact on the wind speed and direction, especially for low-wind situation. Therefore, to further enhance the accuracy of TLTE in highly-risky areas, the wind data can be further downscaled to 100-500m resolution, and refined by considering smaller-grain size terrain information by using tools such as WindNinja [20]. The downscaling of wind data can be selected as a post-processor of weather data to improve the accuracy of TLTE.

In this paper, combining the proposed approximate analytical solution with the NWP data, an efficient system-wide simulation method of TLTE is realized. Our method periodically retrieves environmental data and simulates system-wide TLTE [21] within the time coverage of NWP. With approximate analytical solutions and fast updates of solutions for the current operating condition, the proposed method is over 1000 times faster than conventional methods. Moreover, for the lines with potential risks of over-temperature, we can also realize further analyses, e.g. estimating the risk of over-temperature by considering uncertainty of environmental factors and deriving the permissible time for remedial actions before over-temperature [22]. These results can be visualized in a space on environmental parameters to facilitate analysis and decision support for planning and operation.

In the rest of the paper, Section II derives the approximate analytical solution of TLTE. Section III applies analytical solutions into system-wide analysis of over-temperature events combing NWP data. Section IV verifies the accuracy and practicality of the method on a single conductor model, and Section V demonstrates the accuracy and efficiency of system-wide simulation in NPCC system. Section VI draws conclusions.

II. ANALYTICAL SOLUTIONS OF LINE TEMPERATURE

A. Model of TLTE

Per the IEEE-738 standard [14], the TLTE follows the following differential equation:

$$mC_p \frac{dT_c}{dt} = q_i + q_s - q_c - q_r \quad (1)$$

where $mC_p$ is the heat volume per length of the conductor; $T_c$ is the temperature of conductor and $q_i = I^2 R(T_c)$ is the joule heat generated by current $I$ on a temperature-dependent resistance $R(T_c) = R_0 + \alpha_R (T_c - T_0)$; $q_s$ is the power of heat absorbed from sun light radiation as:

$$q_s = \alpha Q_{se} \sin(\theta) A', \quad \theta = \arccos[\cos(H_c)\cos(Z_c - Z_l)] \quad (2)$$

Here, $\alpha$ is absorptivity coefficient of the conductor; $Q_{se}$ is the solar radiation power per unit area; $A'$ is the projected area of conductor per unit length; $H_c$ is the sun altitude angle, and $Z_c$ and $Z_l$ are azimuth angles of the sun and line.

In (1), $q_r$ is the radiation heat emitted from the conductor.

$$q_r = 17.8\varepsilon D\left[\left(\frac{T_c + 273}{100}\right)^4 - \left(\frac{T_a + 273}{100}\right)^4\right] \text{(W/m)} \quad (3)$$

where, $\varepsilon$ is emissivity coefficient of the conductor; $D$ is the diameter (mm); $T_a$ is ambient temperature (°C).

$q_c$ in (1) is the power of convection heat loss given by:

$$q_c = \max\{q_c^h, q_c^l, q_c^s\} \quad (4)$$

where, $q_c^h$, $q_c^l$ and $q_c^s$ correspond to different wind speeds

$$q_c^h = 0.754 K_a N_R^{0.6} k_f (T_c - T_a) \text{ (W/m)} \quad (5)$$

$$q_c^l = K_a[1.01 + 1.35 N_R^{0.52}] k_f (T_c - T_a) \text{ (W/m)} \quad (6)$$

$$q_c^s = 3.645 \rho_f^{0.5} D^{0.75} (T_c - T_a)^{1.25} \text{ (W/m)} \quad (7)$$

$K_a$ relates to the angle between wind and line $\phi \in [0°, 90°]$

$$K_a = 1.194 - \cos\phi + 0.194\cos 2\phi + 0.368\sin 2\phi \quad (8)$$

and $k_f$ is thermal conductivity of air. $N_R$ is Reynolds number

$$N_R = (D\rho_f V_w) / \mu_f \quad (9)$$

where $V_w$ is wind speed, $\rho_f$ is air density, and $\mu_f$ is air viscosity. $k_f$, $\rho_f$ and $\mu_f$ are dependent on $T_c$ and $T_a$ [14].

B. Approximate analytical solution

Next, we derive the approximate analytical solution of TLTE. Based on (2)-(9), the eq. (1) can be reformulated as

$$\frac{dT_c}{dt} = Q_{si} - \beta(T_c, T_a, I, V_w, \theta_w, H_c, Z_c, Z_l)(T_c - T_a) \quad (10)$$

where $\theta_w$ is wind direction. When the line current and weather condition are given, $\beta$ is a function of conductor temperature:

$$\beta(T_c) = \frac{C_c - I^2 \alpha_R + \frac{17.8\varepsilon D(T_c + T_a + 576)}{10000}\left[\left(\frac{T_c + 273}{100}\right)^2 + \left(\frac{T_a + 273}{100}\right)^2\right]}{mC_p} \quad (11)$$

where term $C_c$ comes from convection heat

$$C_c = \max\{0.754 K_a N_R^{0.6} k_f, K_a(1.01 + 1.35 N_R^{0.52}) k_f, \\ 3.645 \rho_f^{0.5} D^{0.75} (T_c - T_a)^{0.25}\} \quad (12)$$

and the term $Q_{si}$ is

$$Q_{si} = \frac{I^2 R(T_a) + q_s(Q_{se}, H_c, Z_c, Z_l)}{mC_p} \quad (13)$$

The assumption that the $V_w$, $\theta_w$, $T_a$, $H_c$, $Z_c$, $I$ and $Q_{se}$ can be regarded as constants (e.g. average value) over a period is reasonable since the heat volume of the conductor acts as a low-pass filter suppressing the effect of high-frequency fluctuation in parameters. Let $\Delta T = T_c - T_a$ and denote $\beta$ as the function of new variable $\Delta T$, i.e. $\beta_\Delta(\Delta T)$, then (10) is transformed as

$$\frac{d\Delta T}{dt} = Q_{si} - \beta_\Delta(\Delta T)\Delta T \tag{14}$$

Tests show that in most cases, $\beta_\Delta(\Delta T)$ has good linearity with $\Delta T$ as illustrated in Fig. 1, so it can be approximated by

$$\beta_\Delta(\Delta T) = \beta_{\Delta 0} + \beta_{\Delta T}\Delta T \tag{15}$$

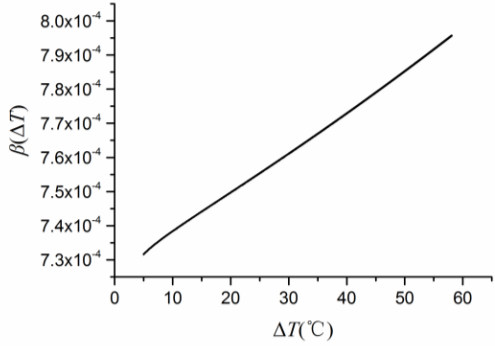

Fig. 1. The linearity of $\beta_\Delta(\Delta T)$ to $\Delta T$. Conductor: Drake (diameter 28.1mm), on July 1st, 12:00PM, at latitude 30°N, line direction E-W, $V_w$=1.3m/s, wind direction E-W, $T_a$=40°C.

Thus, equation (14) is approximated as a constant-coefficient Ricatti equation (16) with solution (17).

$$\frac{d\Delta T}{dt} = Q_{si} - \beta_{\Delta 0}\Delta T - \beta_{\Delta T}\Delta T^2 \tag{16}$$

$$T_c^{Ric}(t) = \frac{\Delta_B - \Delta_A C' e^{-\beta_{\Delta T}(\Delta_A + \Delta_B)t}}{1 + C' e^{-\beta_{\Delta T}(\Delta_A + \Delta_B)t}} + T_a \tag{17}$$

where

$$\Delta_B = T_e - T_a, \Delta_A = \Delta_B + \frac{\beta_{\Delta 0}}{\beta_{\Delta T}}, C' = \frac{T_e - T_{c0}}{\Delta_A + T_{c0} - T_a} \tag{18}$$

Here $T_{c0}$ is initial temperature, and $T_e$ is the steady-state temperature calculated by assuming $d\Delta T/dt=0$ in (16).

Thus, the approximate analytical solution of TLTE is obtained, whose advantages are twofold: 1) the line temperature variation can be calculated efficiently, avoiding directly integrating (1) like conventional numerical methods; 2) the analytical formulation can be conveniently used in system risk assessment.

Moreover, it is desirable to obtain an even simpler form as the solution of a first-order differential equation.

$$T_c^{simp}(t) = T_e + (T_{c0} - T_e)e^{-\beta' t} \tag{19}$$

The aim is to find $\beta'$ so that $T_c^{simp}(t)$ is not lower than $T_c^{Ric}(t)$ and their difference is as small as possible. It is proved that the optimal $\beta'$ is (detailed proof is provided in Appendix A):

$$\beta' = \beta_{\Delta T}(\Delta_A + \Delta_B) = \sqrt{\beta_{\Delta 0}^2 + 4Q_{si}\beta_{\Delta T}} \tag{20}$$

The first-order solution guarantees conservativeness, and the difference between $T_c^{simp}(t)$ and $T_c^{Ric}(t)$ is restricted by

$$0 \leq T_c^{simp}(t) - T_c^{Ric}(t) \leq \frac{(\sqrt{1+C'}-1)^2}{1+C'}(T_e - T_a + \Delta_A) \tag{21}$$

*C. Algorithm for obtaining analytical solution*

To obtain an analytical solution, first we solve the steady-state temperature $T_e$ with the Newton-Raphson (N-R) method as follows. At steady-state temperature, the equality holds:

$$Q_{si} - \beta_\Delta(T_{\Delta e}) \cdot T_{\Delta e} = 0 \tag{22}$$

where $T_{\Delta e} = T_e - T_a$. In most cases, from initial temperature $T_{c0}$ to steady-state temperature $T_e$, $\beta$ does not change much, so we can approximate $\beta_\Delta(T_{\Delta e}) \approx \beta_\Delta(T_{\Delta c 0})$. Then from (14) the steady-state temperature is approximated by (23) leading to a mismatch on the left-hand side of (24), denoted by $\Delta Q$.

$$\hat{T}_{\Delta e} = \frac{Q_{si}}{\beta_\Delta(T_{\Delta c 0})} \tag{23}$$

$$\Delta Q = Q_{si} - \beta_\Delta(\hat{T}_{\Delta e}) \cdot \hat{T}_{\Delta e} \tag{24}$$

Then we have

$$\frac{d\Delta Q}{d\hat{T}_{\Delta e}} = -\beta_\Delta(\hat{T}_{\Delta e}) - \frac{d\beta_\Delta}{d\hat{T}_{\Delta e}} \cdot \hat{T}_{\Delta e} \approx -\beta_\Delta(\hat{T}_{\Delta e}) - \frac{\beta_\Delta(\hat{T}_{\Delta e}) - \beta(T_{\Delta c 0})}{\hat{T}_{\Delta e} - T_{\Delta c 0}}\hat{T}_{\Delta e} \tag{25}$$

The Newton-Raphson correction on $\hat{T}_{\Delta e}$ is

$$\Delta\hat{T}_{\Delta e} = -\Delta Q \Big/ \frac{d\Delta Q}{d\hat{T}_{\Delta e}} \tag{26}$$

Perform steps (24) to (26) repeatedly until the heat power mismatch becomes less than a given threshold, i.e. $\Delta Q < \varepsilon_Q$. Converged $T_{\Delta e}$ derives steady-state temperature $T_e = T_{\Delta e} + T_a$. The other parameters in the analytical solution can be derived:

$$\beta_{\Delta T} = \frac{\beta_\Delta(T_{\Delta e}) - \beta_\Delta(T_{\Delta c 0})}{T_{\Delta e} - T_{\Delta c 0}} \tag{27}$$

$$\beta_{\Delta 0} = \beta_\Delta(T_{\Delta c 0}) - \beta_{\Delta T}(T_{c0} - T_a) \tag{28}$$

Following the above steps to obtain $T_e$, $\beta_{\Delta 0}$ and $\beta_{\Delta T}$, then the analytical solutions are derived based on (17)-(20).

*D. Efficient update of solution when line current changes*

The merit of the approximate analytical solution is also in efficient update of solutions when system states change. Assume the environmental variables $V_w$, $\theta_w$ and $T_a$ to remain the same, and $T_e$, $\beta_{\Delta 0}$, $\beta_{\Delta T}$ and $Q_{si}$ under current $I$ (namely the reference current) have been calculated by (22)-(28). When current changes to $I'$, note that $\beta_{\Delta T} \approx \partial\beta/\partial T$ and $\partial\beta/\partial T$ does not depend on $I$ according to (11), so approximately $\beta_{\Delta T}(I') = \beta_{\Delta T}(I)$, and

$$Q_{si}(I') = \frac{(I'^2 - I^2)R(T_a)}{mCp} \tag{29}$$

$$\beta_{\Delta 0}(I') = \beta_{\Delta 0}(I) + \frac{(I^2 - I'^2)\alpha_R}{mCp} \tag{30}$$

From (22), the steady-state temperature under current $I'$ can be estimated by

$$T_e(I') = \frac{\sqrt{\beta_{\Delta 0}^2(I') + 4\beta_{\Delta T}(I')Q_{si}(I')} - \beta_{\Delta 0}(I')}{2\beta_{\Delta T}(I')} + T_a \tag{31}$$

Then the TLTE under new current $I'$ can be updated with (17) and (19). It indicates that only one N-R iteration is required if the environmental factors remain unchanged. The analytical update of line current will be very useful in batch analysis of multiple operation states (e.g. N-k contingency analysis), in which only line current is changed and the new solutions can directly be updated from (29)-(31).

### III. EVALUATION OF OVER-TEMPERATURE EVENTS

*A. Sources of environmental data*
*1) NWP and historical data*

Calculating the system-wide TLTE requires reliable sources of environmental data, mainly including ambient temperature, wind speed and wind angle. Table I shows data sources and uncertainties in different applications including planning, scheduling and operations. In operations, environmental data can be obtained from the numerical weather prediction (NWP) analysis and forecast results, and the uncertainty mainly comes from the error of the NWP [23]. Major NWP models used in the US are listed in Table II. The scheduling of power systems can vary on a daily or weekly basis, or even on monthly basis. Due to the limitation of NWP spatial resolution and time outreach, the NWP data can only cover up to weekly scheduling, and the error grows with the increase of NWP outreach. Therefore, longer-term scheduling will depend more on historical data, and the environmental data will have larger uncertainties. The historical environmental data can be collected from weather observations and/or the analysis of historical NWP results [24].

TABLE I. ENVIRONMENTAL DATA SOURCE AND UNCERTAINTY

| Operation | Source | NWP |
|---|---|---|
| | Uncertainty | NWP Error |
| | Source | Historical data/NWP |
| | Uncertainty | Historical data distribution/NWP Error |
| | Source | Historical data |
| | Uncertainty | Historical data distribution |

TABLE II. NOAA NWP MODEL SPECIFICATIONS (BY 2016)

| Models | CFS | GFS | NAM | RAP | HRRR |
|---|---|---|---|---|---|
| Coverage | Global | Global | North America | North America | CONUS, Alaska |
| Outreach | 9 mo. | 16 days | 3.5 days | 21 hr. | 18 hr. |
| Refresh | 6 hr. | 6 hr. | 6 hr. | 1 hr. | 1 hr. |
| Time-step | 6 hr. | 3 hr. | 3 hr. | 1 hr. | 15 min. |
| Resolution | ~56 km | ~28 km | ~12km | ~13km | ~3km |

*2) Post-processing of weather data*

The accuracy of simulating TLTE significantly depends on the accuracy of weather data. Due to the complex nature of atmospheric system and the limitation in techniques in meteorological observation and NWP computation, there is inevitably error of NWP results as compared with real events. In addition, the limited spatial resolution of NWP may lead to inaccuracy of TLTE. Therefore, to enhance the quality of weather data and improve the accuracy of TLTE simulation, it is desirable to conduct post-processing of NWP results, e.g. make corrections to reduce the errors or refine the result to higher resolution.

Take the NWP services in North America as an example. The environmental data such as near-surface temperature (2m altitude temperature) and wind speed (10m altitude wind x- and y-component) are available with spatial resolution as high as 3km (HRRR). The operational HRRR has enhanced the accuracy of near-surface prediction by considering terrain factors, yet such spatial resolution in some cases still cannot sufficiently reflect the terrain of smaller scales (e.g. small hills, valleys). Since terrain has non-negligible impacts on environmental factors, especially on wind speeds and directions, more detailed terrain data are helpful to further improve the accuracy of simulation [25]. The downscaling of wind data considering even higher terrain resolution can be achieved with software such as WindNinja [20]. WindNinja is capable of retrieving high-resolution terrain data from online databases and utilizing NWP results to generate terrain-corrected wind vectors at spatial resolution of 100-450m.

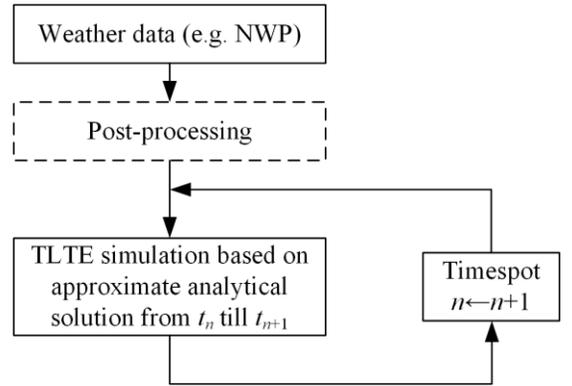

Fig. 2. General procedure of TLTE simulation. The dashed line indicates that the post-processing is not absolutely necessary, but can improve the data quality.

Moreover, some existing techniques can be utilized to reduce the error of NWP results. For example, ref. [26] proposed a statistical method to make corrections to NWP local wind forecast by using historical observations. The test case in Ireland has verified that such technique can effectively reduce the bias and root-mean-square error of NWP forecast results.

In summary, the abovementioned post-processing techniques are exerted after retrieving weather data and before simulating the TLTE (as shown in Fig. 2), which improve the quality of weather data and thus can enhance the practicality of the proposed method.

*B. Simulation of system-wide TLTE*

The proposed analytical solutions can be utilized in TLTE either on normal operating conditions or under contingencies. Take the contiguous US and Alaska as examples. The high-resolution rapid-refresh (HRRR) mode is covered. NWP data is open access to the public and can be retrieved from the NOAA Operational Model Archive and Distribution System (NOMADS). Since the line temperature largely depends on the environmental factors, the transmission line should be divided into segments in the analysis. For the HRRR resolution of 3km, the length of a line segment should not exceed 3km.

To further accelerate computation, the line segments with close positions and similar environmental factors can also be approximately regarded as the same, and then the parameters of analytical solutions can be regarded as the same. This paper also realizes line segment clustering (LSC) with K-means method [27]. Then the analytical solution parameters within a cluster can be regarded as the same and can be computed only once, further reducing computational burden. The parameters of LSC are the geographical coordinates, ambient temperature, wind speed, wind direction, line segment direction, and conductor type.

The efficient system-wide simulation of TLTE can be realized with the proposed method. The NWP data are refreshed hourly. Assume the studied system states (e.g. a set of N-k contingency states) have been obtained. Then for each operational state and each line segment, an analytical temperature trace for up to 18 hours is obtained, as illustrated in Fig. 3. Note that the actual outreach of TLTE prediction will be a bit less than 18 hours. It is because the data assimilation of NWP needs to collect current observations and combine them with forecasts from the previous cycles to produce new forecasts, which generally takes about one hour (indicated by the shadowed areas along the timeline shown in Fig. 3). Therefore, the NWP forecast starting at time $t_0$ will be ready at about an hour after $t_0$, and the real outreach of line temperature traces is about 17 hours. Such time outreach should be enough for online decision support against the overheat of transmission lines.

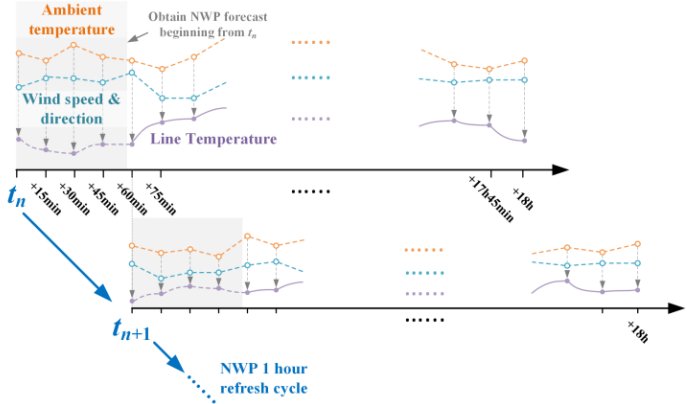

Fig. 3. Computation cycle in online operation.

## C. Estimating the risk of over-temperature
### 1) Over-temperature risk at steady state

The analytical solution can also facilitate further study of lines of interest (e.g. heavy-loaded lines with higher risk of overheat). For example, we can investigate the risk of exceeding a temperature limit $T_{th}$ considering uncertainties in environmental data. The steady-state temperature $T_e$ is determined by the ambient temperature $T_a$, wind speed $V_w$ and direction angle $\theta_w$. Given $T_a$ and $\theta_w$, the wind speed corresponding to steady-state temperature reaching $T_{th}$ can be uniquely determined. When $T_e = T_{th}$, the convection heat is determined from (1):

$$q_c(T_{th}, T_a, V_w, \theta_w) = I^2 R(T_{th}) + q_s - q_r(T_{th}, T_a) \quad (32)$$

If $q_c(T_{th}, T_a, V_w, \theta_w) < q_c^s$ then $T_{th}$ cannot be reached. Otherwise, we can intersect two wind speeds from $q_c^l$ and $q_c^h$:

$$V_w^l = \left( \frac{q_c(T_{th}, T_a, V_w, \theta_w)}{1.35 K_{angle} k_f (T_{th} - T_a)} - \frac{1.01}{1.35} \right)^{\frac{1}{0.52}} \frac{\mu_f}{D_0 \rho_f}$$

$$V_w^h = \left( \frac{q_c(T_{th}, T_a, V_w, \theta_w)}{0.754 K_{angle} k_f (T_{th} - T_a)} \right)^{\frac{1}{0.6}} \frac{\mu_f}{D_0 \rho_f} \quad (33)$$

Then the smaller one of them, i.e., $V_w^{th} = \min\{V_w^h, V_w^l\}$, is the estimated maximum wind speed to cause over-temperature.

Since $V_w^{th}$ corresponds to $T_{th}$, any wind speed lower than $V_w^{th}$ will result in steady-state temperature higher than $T_{th}$. Hence, $V_w \leq V_w^{th}$ is the region of potential over-temperature events in the environmental parameter space. Given the marginal distribution of $V_w^{th}$, the over-temperature probability can be estimated by

$$\Pr\{T_e \geq T_{th} \mid T_a, \theta_w\} = \Pr\{V_w \leq V_w^{th} \mid T_a, \theta_w\} = F_{V_w \mid T_a, \theta_w}(V_w^{th}) \quad (34)$$

where $F_{V_w \mid T_a, \theta_w}(\cdot)$ is the marginal CDF of wind speed.

In the environmental parameter space $S_p$, the overall probability that steady-state temperature exceeds limit can be estimated by integration over the space of $\theta_w$ and $T_a$.

$$\Pr\{T_e \geq T_{th}\} = \int_{S_p} F_{V_w \mid T_a, \theta_w}(V_w^{th}) p(T_a, \theta_w) d\Omega$$
$$\doteq \sum_{T_a, \theta_w} F_{V_w \mid T_a, \theta_w}(V_w^{th}) p(T_a, \theta_w) \Delta T_a \Delta \theta_w \quad (35)$$

### 2) Assessing the time to over-temperature

Besides the risk that steady-state line temperature exceeds the limit, the transient as well as the time for the line temperature to rise to the limit are also desirable because it indicates the time left for the operators to take remedial actions to relieve overheat. Given $T_a$ and $\theta_w$, and any $T_e$ between the maximum possible steady-state temperature $T_{e\max}$ (obtained from the possible minimum wind speed $V_{w\min}$) and $T_{th}$, the time $t$ for reaching $T_{th}$ is calculated by the following procedure:

**Step 1.** Calculate $\beta_{\Delta e} = Q_{si} / (T_e - T_a)$.

**Step 2.** Calculate convection heat at the steady state.
$$q_c(T_e, T_a, V_w, \theta_w) = I^2 R(T_e) + q_s - q_r(T_e, T_a)$$

**Step 3.** Calculate $V_w^{th}$ from $q_c(T_e, T_a, V_w, \theta_w)$ with (24).

**Step 4.** Calculate $\beta(T_{c0})$, $\beta_{\Delta T} = (\beta_{\Delta e} - \beta(T_{c0})) / (T_e - T_{c0})$, $\beta_{\Delta 0} = \beta(T_{c0}) - \beta_{\Delta T}(T_{c0} - T_a)$, and $\beta' = \sqrt{\beta_{\Delta 0}^2 + 4Q_{si}\beta_{\Delta T}}$

**Step 5.** Calculate time to $T_{th}$ with the first-order solution (19):

$$t = \frac{1}{\beta'} \ln \frac{T_e - T_{c0}}{T_e - T_{th}}$$

Or one can derive it from the second-order solution (17):

$$t = \frac{1}{\beta'} \ln \left( \frac{T_e - T_{c0}}{T_e - T_{th}} \frac{T_{th} + T_e + \beta_{\Delta 0}/\beta_{\Delta T} - 2T_a}{T_{c0} + T_e + \beta_{\Delta 0}/\beta_{\Delta T} - 2T_a} \right)$$

All the points of $t$ form a region indicating how much time is left for the conductor to reach $T_{th}$ in the space of environmental parameters. Such results can be conveniently visualized and thus can enhance situational awareness of transmission lines and facilitate decision support against overheat.

## IV. TESTS OF THE CONDUCTOR MODEL
### A. Verification of approximate analytical solutions

Firstly, we test the derivation of steady-state temperature by using the N-R iteration described by (22)-(26). We set 50000 instances with various parameters: line diameters vary from 0.5cm to 4.75cm (covering all standard ACSRs); initial temperature from 20°C to 100°C; ambient temperature from 0°C to 40°C; line current from 0 to 200% of the nominal current; wind speeds from 0 to 10m/s. All the cases with steady-state temperatures below 300°C converge with respect to a tolerance $\varepsilon_Q = 10^{-6}$ W/m. Over 95% of cases converge within 10 iterations (Fig. 4). All the 50000 instances only cost 1.166s time in total.

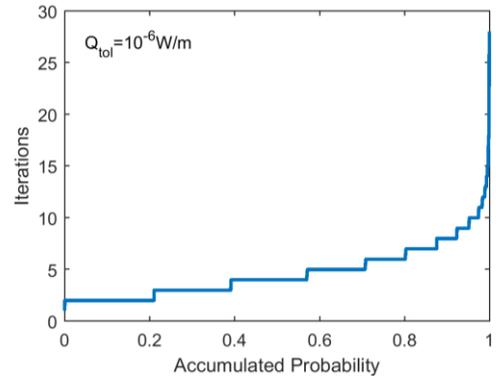

Fig. 4. Steady state temperature iteration times.

TABLE III. PARAMETERS FOR TEMPERATURE VARIATION CALCULATION

| Parameter | Value |
| --- | --- |
| Wind speed | 0.8m/s |
| Wind direction | 90°(East) |
| Latitude | 30°N |
| Date | July 1st |
| Time | 12:00am |
| Line direction | 90°(W-E) |

| Ambient temperature | 40°C |
|---|---|
| Initial temperature | 50°C |

Then we test the accuracy of analytical solutions on a typical model of transmission line conductor. The tested conductor is ACSR "Drake" [14]. First we demonstrate the results of TLTE calculation by comparing different analytical solutions with the results obtained by the numerical method. Related parameters are listed in Table III.

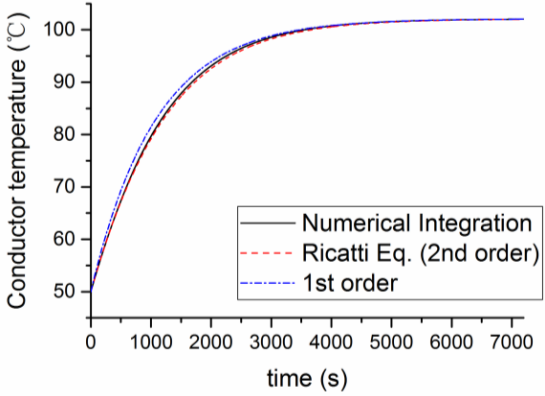

Fig. 5. Comparison of analytical solutions with numerical results.

Set the current as $I = 800$A and compare the solutions given by conventional numerical integration, the solution of Ricatti equation (17) and the first-order solution (19), as shown in Fig. 5. Both analytical solutions match well with the numerical result. The errors of analytical solutions compared with numerical integration solution are presented in Table IV. The Ricatti equation solution gives negative errors, which means slightly lower temperature than numerical results (i.e. $\delta T^-$) and time lag of reaching a certain temperature (i.e. $\delta t^-$). While first order equation can guarantee conservativeness of results, which almost only gives limited positive error compared with numerical results. The accuracy and conservativeness of the first-order solution is desirable for security analysis. With a satisfactory accuracy, a simpler form than the Ricatti equation solution, and a higher efficiency than the numerical method, the first-order analytical solution is appealing in online simulation of TLTE and promising for a higher-level risk assessment.

TABLE IV. ERRORS OF ANALYTICAL SOLUTIONS

|  | Ricatti eq. solution | First-order solution |
|---|---|---|
| max$\delta T^+$ (°C) | 0 | 1.9938 |
| max$\delta t^+$ (s) | 0 | 72.8 |
| max$\delta T^-$ (°C) | **0.5548** | **0.0017** |
| max$\delta t^-$ (s) | **42.2** | **23.5** |

*B. Analytical solutions by updating line current*

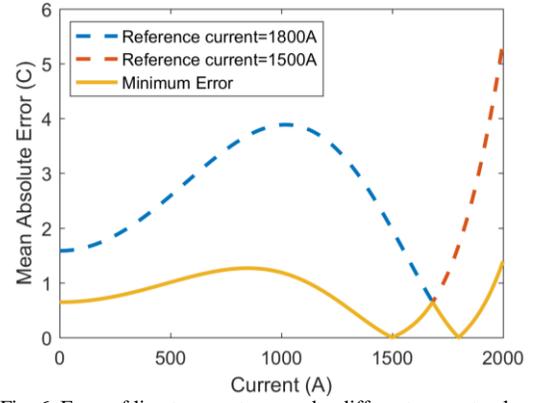

Fig. 6. Error of line temperatures under different current values.

With an analytical solution in which the parameters are obtained from the N-R iterations of (22)-(26), solutions at other line current values can be approximated from (29)-(31). Fig. 6 demonstrates the accuracy of simulating TLTE under different currents. The suggested ampere rating of the conductor is around 1000A. There are two curves corresponding to reference currents as 1500A and 1800A, respectively. From the result, before the current reaches the 1500A reference, the error is maintained below 1.5°C, while beyond 1500A, the error rises sharply. So another reference current is needed if the current can reach 2000A. With supplemented reference current as 1800A, the error in the 0-2000A range is limited under 1.5°C. Extensive tests on various conductor models show that within 0-200% loading region, at most two reference current values are required to limit the error under 2°C. Since normally the protection setting point is around the 200% loading level, this method can generally cover the common long-term operating states of transmission lines. Because updating current values avoids re-calculation of N-R iterations, the computational efficiency is greatly enhanced.

*C. Over-temperature region of sample line segment*

Considering the uncertainty of the environmental factors, the probability of steady-state over-temperature events can be evaluated as discussed in III.C. The wind is assumed to follow Weibull distribution, as shown in the wind rose in Fig. 7. The ambient temperature is assumed to follow uniform distribution between 30°C and 40°C.

The binning of probability distribution of environmental factors affects both the accuracy and computational efficiency. In this case, 2-D binning of ambient temperature and wind direction is needed. Table V demonstrates the estimated probability and computation time under different binning densities. It shows that even with the coarsest 25×25 binning, the result still does not deviate much (around 3%) from the 500×500 binning. The results verify that the probability of over-temperature can be efficiently estimated with satisfactory accuracy. The method can be utilized for risk assessment in both planning and operations.

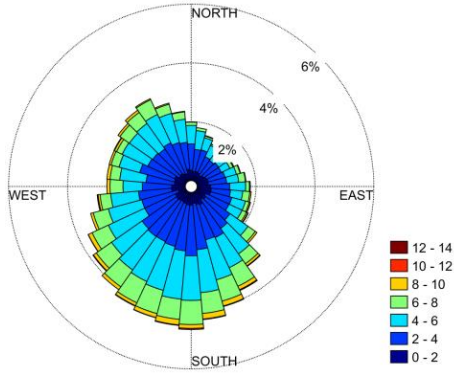

Fig. 7. An example wind rose.

TABLE V. PROBABILITY OF OVER-TEMPERATURE UNDER DIFFERENT BINNING

| Binning | Probability | Error (%) | Time (s) |
|---|---|---|---|
| 500×500 | 0.073696 | --- | 12.2639 |
| 200×200 | 0.073592 | 0.141 | 2.0263 |
| 100×100 | 0.073355 | 0.463 | 0.55621 |
| 50×50 | 0.072623 | 1.456 | 0.14911 |
| 25×25 | 0.071291 | 3.263 | 0.03692 |

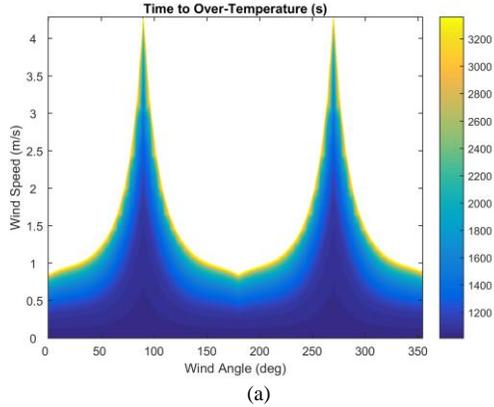

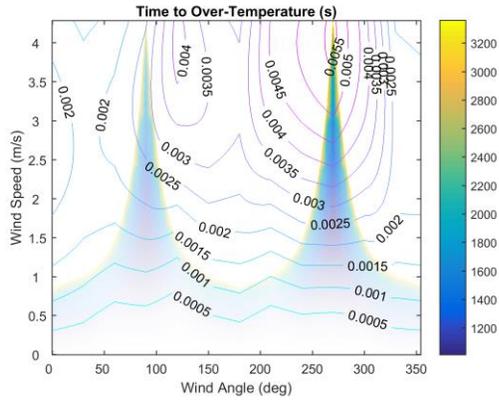

Fig. 8. Time to over-temperature regions.

The line segments with relatively high over-temperature probability can be further analyzed, e.g. the time to over-temperature characteristics in environmental factor space. An example of the time-to-over-temperature region is illustrated in Fig. 8a, with x- and y-axes as wind direction and wind speed. The region is calculated efficiently following the methods in III.C.2 (only 1/4 of the region needs computation due to the symmetry in wind direction). Such a region can conveniently indicate the time left before the conductor temperature reaches the limit, i.e. the time for taking control actions to relieve overheat before it occurs.

The time-to-over-temperature region can also be combined with the probability distribution of environmental factors, as shown in Fig. 8b. The contour plot masked over the region is the probability density of corresponding wind direction and speed. In Fig. 8b transparency is masked over the region to highlight the more possible environmental parameters as well as the corresponding time-to-over-temperature characteristics.

## V. APPLICATION IN NPCC SYSTEM

### A. TLTE simulation with NWP data

We test the method on the NPCC 140-bus, 233-line system located in the northeast of the US and Canada. In this case, we assume the types of the conductors are set as in Table VI.

TABLE VI. TYPICAL TRANSMISSION LINE CONDUCTOR SPECIFICATION [6]

| Voltage | Type | bundle | Diameter(mm) |
|---|---|---|---|
| 765kV | Pheasant ACSR | 4 | 35.10 |
| 500kV | Bittern ACSR | 3 | 34.16 |
| 345kV | Cardinal ACSR | 2 | 30.38 |
| 230kV | Drake ACSR | 2 | 28.14 |
| 138kV | Ibis ACSR | 1 | 19.89 |

The example near-ground temperature and wind vector distribution in NPCC area is demonstrated in Figs. 9 and 10.

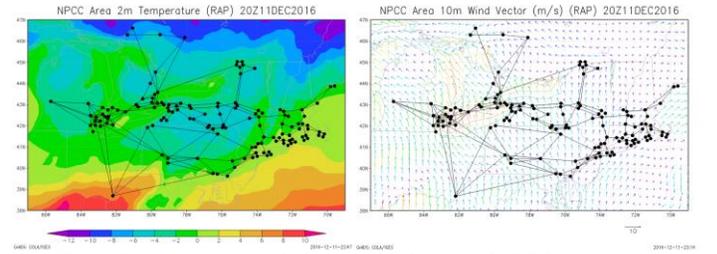

Fig. 9. Temperature and wind vector distribution of NPCC area (RAP).

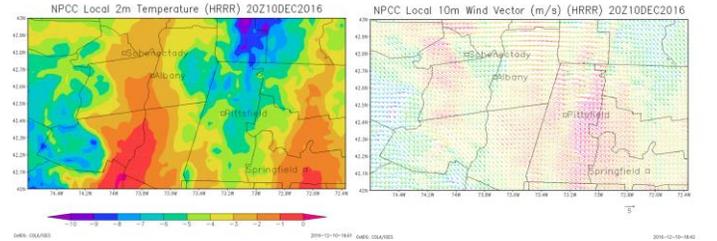

Fig. 10. Temperature and wind vector distribution of partial area (HRRR).

The total estimated length of transmission lines in the NPCC system is 19,607 km. Using the 3-km HRRR data, the whole system will be cut into more than 6000 segments. In this case, to demonstrate the efficiency of the proposed method, we choose an even smaller segment length of no more than 1km, and all the transmission lines are divided into $N_L$=19953 line segments.

The HRRR results with starting hour $t_0$ contain forecasts of every 15 minutes until hour $t_0 + 18$, i.e. 73 time points in total. The traditional method does numerical integration every 15 minutes consecutively for each line segment, and in every studied system operation state. While based on the proposed method using an analytical solution, for each line segment and each time point, analytical solution parameters are calculated no more than twice (only once in most cases). And then temperature evolutions under all studied operation states are obtained efficiently by updating the current values on line segments, which takes much less time. The proposed method is particularly advantageous in the analysis for multiple operation states, e.g. N-k contingency analysis. If the traditional method costs time $\tau_i$ for one system state on average, then for $N_s$ system states, the total time consumption is $N_s\tau_i$. In the proposed method, assume the time for generating analytical solution parameters on all the segments and all the time points to be $\tau_{gp}$, and the average time for obtaining TLTE across the system under one system state to be $\tau_{gs}$, then the total time consumption is $\tau_{gp}+N_s\tau_{gs}$.

The methods are developed and tested in MATLAB on a computer with the Intel Core i7-6700 CPU and 16GB DDR4 RAM. Set the time step for TLTE solution as 5s. We screened a set of 2500 N-k (k≤4) contingencies with a simplified version of the Markovian tree model [28], and then arbitrarily selected 10 contingencies from the set to test the average computational efficiency. It is tested in this case that $\tau_i=5540s$, $\tau_{gp}=62.2s$ and $\tau_{gs}=3.45s$. Since $\tau_{gs}\ll\tau_i$, the proposed method has much higher efficiency, especially for N-k contingency analysis.

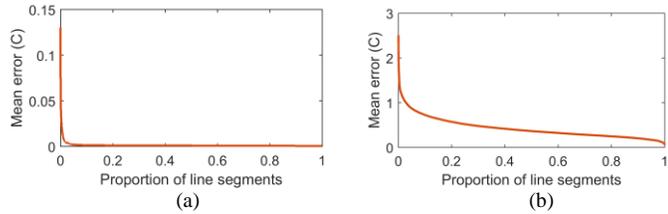

Fig. 11. Mean error of TLTE. (a) w/o line segment clustering (LSC), (b) w/ LSC.

As for the accuracy of the analytical solutions, all the above TLTE from the proposed method are compared with solutions of conventional methods. The mean error of line segment temperature traces is shown in Fig. 11a. For all the line segments, the errors are below 0.15°C and most errors are nearly 0. This result verifies the satisfactory accuracy of the proposed method.

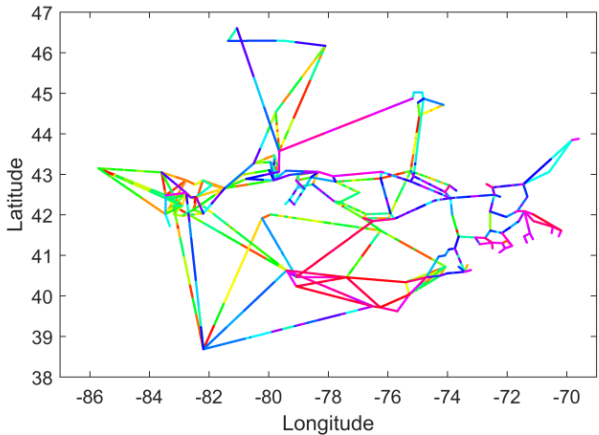

Fig. 12. K-means line segment clustering (LSC) results.

It is tested that in the NPCC system, selecting cluster number $k=500$ can limit the differences within a cluster of ambient temperature under 2°C, wind speed below 2m/s, and wind direction within 10º angle. The clusters of the line segments are shown in Fig. 12. Based on test results, the error of TLTE with LSC is larger than the solutions without LSC, as is shown in Fig. 11b. But for most line segments, the error of line temperature is below 1°C.

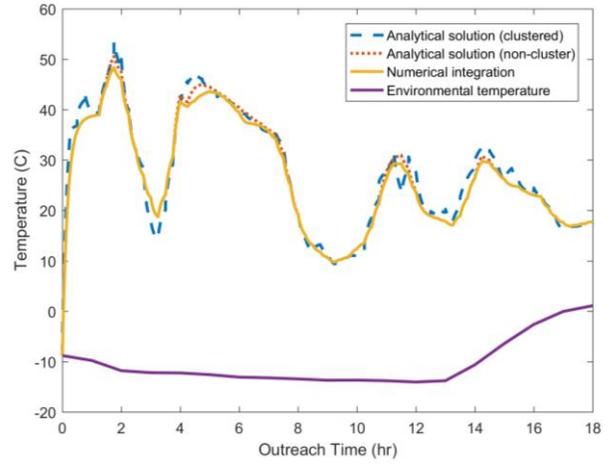

Fig. 13. Comparison of TLTE solutions

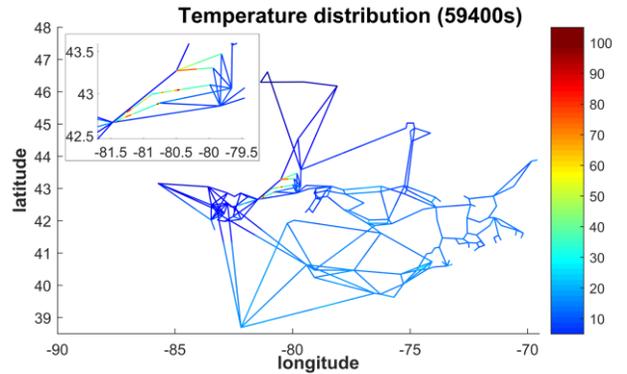

Fig. 14. Line temperature distribution of NPCC system. Note the high-temperature line segments at around 43°N, 80°W.

Fig. 13 compares TLTE solutions from different methods, and the line segment is arbitrarily chosen from the system. It shows that the analytical solutions can accurately simulate the TLTE of the studied line segment. The analytical solution obtained without LSC matches well with the trace from numerical integration in most time spots. The solution with LSC has larger errors, but it matches well with the numerical integration result.

Fig. 14 shows the snapshot of distribution of line segment temperature after lines 67 and 161 quit from operations. With the help of this plot, it is easy for system operators to spot possible overheat line segments at any system operation states.

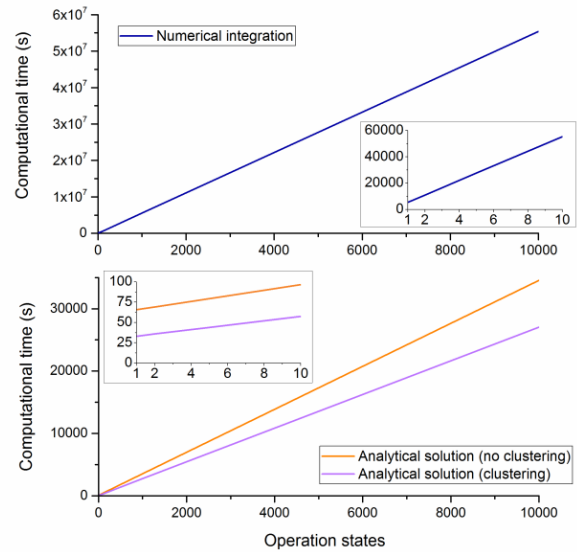

Fig. 15. Computation time of traditional and proposed methods.

As for the computational efficiency of the proposed method with LSC, since at each time spot, only $k$ instead of $N_L$ sets of analytical solution parameters are calculated, the corresponding computation time is significantly reduced to $\tau'_{gp} = 2.43$s. The average time for obtaining analytical solutions is also reduced to $\tau'_{gs} = 2.7$s. The efficiency enhancement of $\tau'_{gs}$ is not as much as $\tau'_{gp}$ because of some fixed computational overheads for, e.g., allocating memory. The LSC improves efficiency of both calculating solution parameters and generating final solutions. However, another big chunk of time consumption is the clustering itself, i.e. $\tau'_c = 28.06$s. The total time cost for line segment clustering is $\tau'_c + \tau'_{gp} + N_s \tau'_{gs}$. Compared with the method without clustering, the time reduction is significant, but not as much as the extent of $k/N_L$ due to the time cost on clustering and fixed overhead. The estimated computational time of traditional numerical integration and the proposed analytical solutions are shown in Fig. 15. The efficiency of the proposed method is significantly higher than the traditional method, particularly when used for the analysis of multiple operation states. Moreover, since the time for obtaining analytical solutions $\tau_{gs}$ and $\tau'_{gs}$ are only around 2-3s, it is fast enough to generate solutions at the time of demand in applications.

Moreover, for fast screening of over-temperature events, we can just calculate temperature at 15-min step on the 73 time points using analytical solutions, which significantly reduces $\tau_{gs}$ and $\tau'_{gs}$. In this case, computing all 2500 N-k scenarios takes only 135.5s in total without LSC, and 80.9s with LSC. The proposed method performs much better than the conventional method (which is estimated to take about $1.3 \times 10^7$s) and hence is promising for online applications.

*B. Refined TLTE results by using downscaled wind data*

As stated previously, the spatial resolution of HRRR mode is 3km. Such resolution is generally sufficient for the system-wide scanning of potential over-temperature area and events, but may be still too coarse for practical alarming and locating the threats. The downscaling of wind data considering terrain effects could be achieved by using tools such as WindNinja, which can downscale the spatial resolution down to 100-450m. Here we adopt a two-stage approach of analyzing the TLTE of the system: first using the 3-km NWP data to scan the TLTE of the whole system, and then pick out the areas with over-temperature events to obtain downscaled wind data and simulate locally refined TLTE.

Take NPCC system as an example, the resolution of refined wind data by WindNinja in Canada is about 400m (~1300ft), and the resolution in the US is 130m-250m (~430-~820ft), which is generally in the same scale of transmission line spans.

First, we analyze the area with high temperature in Fig. 14, select a 262 km × 122km region for wind data downscaling and TLTE refinement. Since the studied region is in Canada, the WindNinja produces downscaled wind data at spatial resolution of 400m (as shown in Figs. 16-17), which is much higher than the resolution of NWP data (Fig. 18). In addition, the downscaling of wind data considers the smaller-grain terrain effect, which is expected to enhance the accuracy of TLTE simulation.

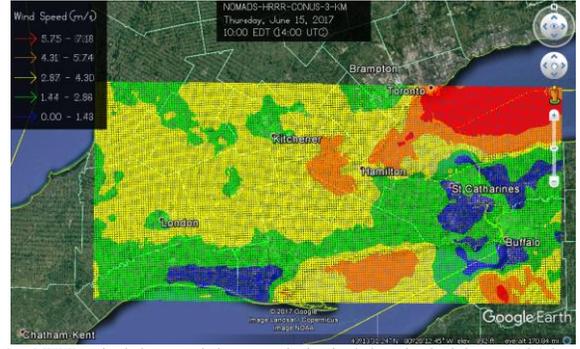
Fig. 16. NWP wind data and downscaled wind data in 262 km × 121 km area.

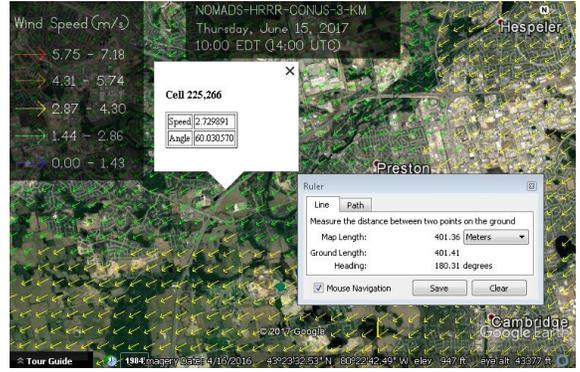
Fig. 17. Zoom-in view of downscaled wind data. The balloon shows the wind speed and direction of the selected data point. The ruler tool shows that the spatial resolution of downscaled wind data is about 400m (~1300ft).

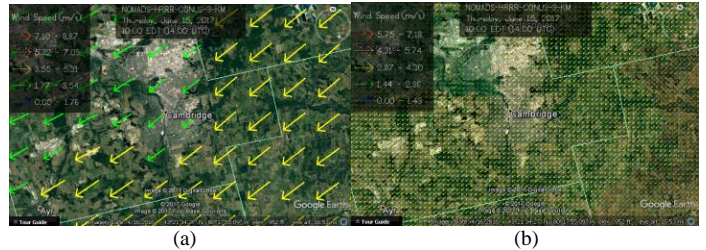
(a)                                     (b)
Fig. 18. Comparison of spatial resolutions of (a) 3-km NWP data and (b) downscaled wind data. The viewpoints from Google Earth are identical.

The TLTE result obtained by using the downscaled wind data is shown in Fig. 20. Compared with the result derived by NWP data (shown in Fig. 19), the refined result gives much smaller grain size of TLTE, which could facilitate people to better locate the potential over-temperature risks.

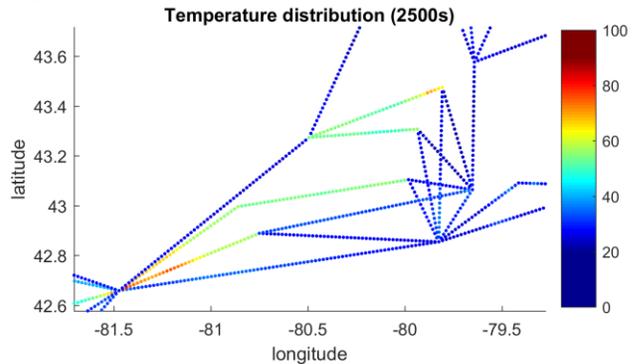
Fig. 19. Local TLTE result by using NWP wind data.

Also comparing the results in Figs. 19 and 20, we can observe that only the lines with high temperature obtained from 3-km NWP data show significantly different temperature distributed along the line after TLTE refinement. While the low-temperature lines do not demonstrate significant temperature difference after refinement, which means the refinement is not very necessary. Therefore, the TLTE refinement with downscaled wind data can be conducted merely on the high-temperature lines rather than refining the whole system, which saves computation resources. The difference between Figs. 19 and 20 is mainly caused by the difference of environmental data, which reflects the significance of accurate environmental data source for realizing practical application of the proposed method. The accuracy of environmental data depends on the accuracy of NWP technique and terrain-correction tools. So the influence of environmental data error on the TLTE accuracy should be investigated in the future. With the improving NWP accuracy [29] and the development of relative techniques [30], the monitoring, analysis and precaution for hazardous environmental-related events in power systems has promising potentials in real applications.

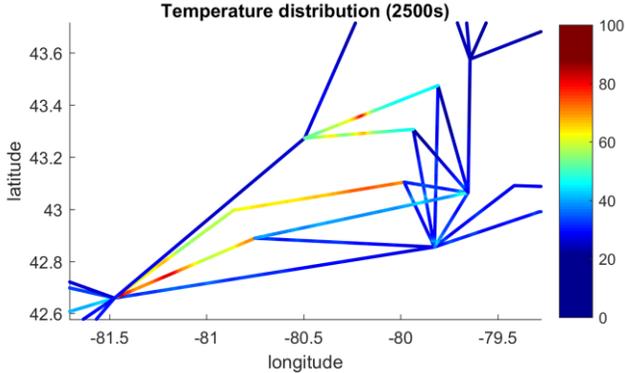

Fig. 20. Refined local TLTE result by using downscaled wind data.

As for the computational efficiency, it has been mentioned in Section V.A that the fast screening of over-temperature events under 2500 N-k contingencies can be finished in 3 minutes. Since the spatial resolution is around 10 times that of 3-km HRRR on average, so even if the whole NPCC system is refined, the computation of 2500 N-k contingencies can be finished within 30 minutes. While the computation of local refinement is much faster: the computation of TLTE for region in Fig. 20 only takes 14.2s.

## VI. Conclusions

This paper proposed an efficient method for the simulation of transmission-line temperature evolution (TLTE). Approximate analytical solutions of TLTE were proposed, which significantly enhance the efficiency over existing methods based on numerical integration. Moreover, this paper proposed fast update of analytical solution when line current changes, further improving performance of batch analysis under multiple operation states. Analytical solutions can also derive over-temperature risk of transmission lines and the time to over-temperature events considering uncertainties of environmental parameters. These results can be conveniently visualized and utilized in planning and operation. Test on a typical conductor model shows analytical solution matches well with numerical integration results, and fast update of line current method efficiently generate solutions with temperature error under 2°C.

Currently, the numerical weather prediction (NWP) provides forecasted environmental data in high spatial and temporal resolution, and with sufficient time outreach for online analysis of transmission lines. With the proposed analytical solution and NWP data, an efficient simulation method of system-wide TLTE is proposed. For the state-of-the-art NWP service covering the US, the time outreach is up to 18 hours and the time step is 15 minutes. The test on the NPCC system with 2500 contingencies shows that the proposed analytical solution based method is thousands of times faster than numerical integration, and system-wide TLTE can be finished in 3 minutes, which is promising for online applications. Moreover, since wind is the most significant factor influencing the TLTE, the NWP wind data can be post-processed (e.g. downscaled considering smaller-grain terrain) before feeding into the analytical solutions, which further improves the accuracy of TLTE in practice.

Furthermore, the analytical solutions together with NWP can be utilized in methods for monitoring and security analysis of transmission systems, which constitutes our future research work.

## Appendix
### A. Derivation of the first-order solution

From (17)-(19), and denote $\beta = \beta_{\Delta T}(\Delta_A + \Delta_B)$, the difference between the first-order solution and the second-order solution is

$$\Delta(t) = T_c^{simp}(t) - T_c^{Ric}(t) = \frac{(\Delta_B + \Delta_A)C'e^{-\beta t}}{1 + C'e^{-\beta t}} - \frac{(\Delta_B + \Delta_A)C'}{1 + C'}e^{-\beta' t} \quad (a1)$$

Note that $\Delta_A + \Delta_B > 0$, and when $C' > 0$, extracting $(\Delta_B + \Delta_A)C'$ does not change the sign of $\Delta(t)$. Let

$$\delta(t) = \frac{\Delta(t)}{(\Delta_B + \Delta_A)C'} = e^{-\beta t}\left(\frac{1}{1 + C'e^{-\beta t}} - \frac{e^{-(\beta' - \beta)t}}{1 + C'}\right) \quad (a2)$$

The conservativeness of first-order solution requires $\delta(t) \geq 0$ when $t \geq 0$. This leads to $\beta' \geq \beta$, because if $\beta' < \beta$, since $1 + C'e^{-\beta t} > 1$, when $t > \ln(1 + C')/(\beta - \beta')$, we have $\delta(t) < 0$.

The derivative of $\delta(t)$ is

$$\delta'(t) = \frac{-\beta e^{-\beta t}}{(1 + C'e^{-\beta t})^2} + \frac{\beta' e^{-\beta' t}}{1 + C'} = e^{-\beta t}\left[\frac{-\beta}{(1 + C'e^{-\beta t})^2} + \frac{\beta' e^{-(\beta' - \beta)t}}{1 + C'}\right] \quad (a3)$$

Since $\beta' \geq \beta > 0$, $\delta'(t)$ monotonically decreases with $t$ ($\delta''(t) < 0$). When $t = 0$,

$$\delta'(0) = \frac{\beta'}{1 + C'} - \frac{\beta}{(1 + C')^2} > \frac{\beta' - \beta}{1 + C'} \geq 0 \quad (a4)$$

and with sufficiently large $t$, $\delta'(t) < 0$. So $\exists t_s \in [0, +\infty)$, s.t. $\delta'(t_s) = 0$. As $\delta''(t) < 0$, $\delta(t_s)$ is the maximum value of $\delta(t)$, i.e. $\Delta(t_s)$ is the maximum difference between first-order and second-order solutions. To study the optimal value of $\beta'$, regard $\delta$ as the function of both $t$ and $\beta'$, and with $\left.\frac{\partial \delta}{\partial t}\right|_{t=t_s} = 0$,

$$\left.\frac{d\delta}{d\beta'}\right|_{t=t_s} = \left.\left(\frac{\partial \delta}{\partial \beta'} + \frac{\partial \delta}{\partial t}\frac{dt}{d\beta'}\right)\right|_{t=t_s} = \left.\frac{\partial \delta}{\partial \beta'}\right|_{t=t_s} = \frac{t_s e^{-\beta' t_s}}{1 + C'} > 0 \quad (a5)$$

Therefore, the smaller $\beta'$, the smaller is the error of first-order solution. Since $\beta' \geq \beta$, and it is easy to verify that $\delta(t) \geq 0$ when $\beta' = \beta$, so we get (20). In this case, the error bound (21) can also be derived by assigning $\beta' = \beta$ to (a2)-(a3) and solving $t_s$ with $\delta'(t_s) = 0$. The derivation for the case of $C' < 0$ is similar.

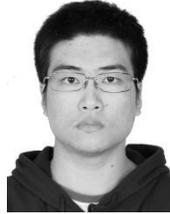

**Rui Yao (S'12-M'17)** received the B.S. degree (with distinction) in 2011 and Ph.D. degree in 2016 in electrical engineering at Tsinghua University, Beijing, China. He was a visiting scholar at EECS, UTK from 2014 to 2015, and was a visiting scholar at the Skolkovo institute of science and technology (Skoltech) in Russia in 2015.

He is currently a postdoctoral research associate at EECS, the University of Tennessee, Knoxville. His research interests include cascading outages, environmental influence on power systems, power system stability, etc.

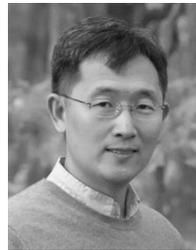

**Kai Sun** (**M'06–SM'13**) received the B.S. degree in automation in 1999 and the Ph.D. degree in control science and engineering in 2004 from Tsinghua University, Beijing, China. He is currently an associate professor at the Department of EECS, University of Tennessee in Knoxville. Dr. Sun is an editor of IEEE Transactions on Smart Grid and an associate editor of IET Generation, Transmission and Distribution. He was a project manager in grid operations and planning at EPRI, Palo Alto, CA from 2007 to 2012. His research interests include dynamics, stability and control of power systems and other complex network systems.

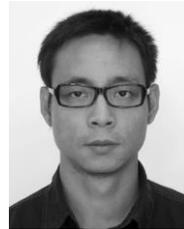

**Feng Liu (M'10)** received his B.S. and Ph.D. degrees in electrical engineering from Department of Electrical Engineering at Tsinghua University, Beijing, China, in 1999, and 2004, respectively.

He is currently an associate professor at the Department of Electrical Engineering of Tsinghua University. His research interests include power system stability analysis and control, robust dispatching, and complex system theory.

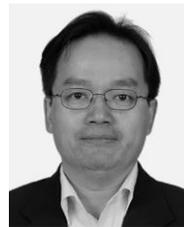

**Shengwei Mei (SM'05-F'15)** received the B.S. degree in mathematics from Xinjiang University, Urumuqi, China, the M.S. degree in operations research from Tsinghua University, Beijing, China, and the Ph.D. degree in automatic control from the Chinese Academy of Sciences, Beijing, China, in 1984, 1989, and 1996, respectively.

He is currently a professor at the Department of Electrical Engineering of Tsinghua University. His research interests include power system analysis and control, robust control and complex systems.